# Tribocorrosion under galvanic interaction of Ti6Al4V and NiCr implant alloys

Camila Dias dos Reis Barros Ph.D.[a], Janaina Cardozo Rocha Ph.D.[b], Ivan Napoleão Bastos Ph.D.[c], José Antônio da Cunha Ponciano Gomes Ph.D.[d]

[a] Ph.D., LabCorr (Corrosion Laboratory), Department of Materials and Metallurgical Engineering (COPPE/UFRJ). Federal University of Rio de Janeiro, Department of Materials and Metallurgical Engineering (COPPE/UFRJ), Av. Horácio de Macedo, 2030 Bloco I Cidade Universitária, Rio de Janeiro, Brazil, zip code: 21941-972, creis@metalmat.ufrj.br.

[b] Research Staff, LabCorr (Corrosion Laboratory), Department of Materials and Metallurgical Engineering (COPPE/UFRJ). Federal University of Rio de Janeiro, Department of Materials and Metallurgical Engineering (COPPE/UFRJ), Av. Horácio de Macedo, 2030 Bloco I Cidade Universitária, Rio de Janeiro, Brazil, zip code: 21941-972, janaina@metalmat.ufrj.br

[c] Professor, Department of Materials Engineering (IPRJ/UERJ), Rio de Janeiro State University, Rua Bonfim, 55 UERJ, Nova Friburgo, RJ, Brazil, zip code 28625-570, inbastos@iprj.uerj.br

[d] Professor, LabCorr (Corrosion Laboratory), Department of Materials and Metallurgical Engineering (COPPE/UFRJ). Federal University of Rio de Janeiro, Department of Materials and Metallurgical Engineering (COPPE/UFRJ), Av. Horácio de Macedo, 2030 Bloco I Cidade Universitária, Rio de Janeiro, Brazil, zip code: 21941-972, ponciano@metalmat.ufrj.br

**Corresponding author:** ponciano@metalmat.ufrj.br

Av. Horácio de Macedo, 2030 Bloco I Cidade Universitária, Rio de Janeiro, Brazil, zip code: 21941-972.

**Abstract**

Micromovements that occur in the joint between dental prostheses and implants can lead to wear-induced degradation. This process can be enhanced by corrosion in the oral environment influenced by the presence of solutions containing fluoride. Moreover, the eventual galvanic interactions between NiCr and Ti alloys can accelerate the wear-corrosion process. In this work, the tribocorrosion process of Ti6Al4V and NiCr alloys used in dental implant rehabilitations immersed in fluoride solutions at different pH values was investigated. The galvanic interaction effect between the alloys was also assessed. Tribocorrosion tests in corrosive media were performed with isolated Ti6Al4V and NiCr alloys, followed by testing with both alloys in contact. The media selected were based on fluoride concentrations and pH values that are possible to be found in oral environments. Analysis of the surfaces after the tribocorrosion tests was carried out using confocal laser microscopy. The wear profile and volume losses were determined by confocal measurements. It was concluded that the galvanic interaction between the alloys increased the tribocorrosion resistance of Ti6Al4V, compared with that of the isolated Ti6Al4V alloy. Ti6Al4V coupled with NiCr reduced the electrochemical potential decay during sliding. The increased resistance was explained by the electrochemical shift of the Ti6Al4V potential from active dissolution to the passive domain.

1. Introduction

Implant fixture, screw-retained abutment, and crown prostheses are components of a dental implant rehabilitation system. The crown is placed over the abutment component. The joint of abutment-implant attach consists of a male-female docking system with the implant fixture, held by screwing. The long success of dental implant rehabilitation depends on proper osseointegration and precise fitting of the joint formed between the crown and the implant. The screw-retained tightening supports the precise fitting and, together with the maintenance of joint settlement, are basic requisites to the success of rehabilitation [1-3]. Various metallic alloys can be used as the components of this dental rehabilitation. The implants fixture and screw are commonly made with Ti alloys, while the abutment uses Ti or casting alloys as NiCr. The abutments can be pre-manufactured or castable. The preference for castable abutments is associated with subgingival prosthesis, a single tooth, and limitation of the prosthetic space available for rehabilitation. The low cost, ease of casting and finishing, adequate mechanical properties, accuracy of fitting, and good adhesion of ceramics make NiCr alloys attractive for castable abutments in implant rehabilitation [2,4-6]. The choice for titanium-based alloys for dental implant fixture is because they have adequate mechanical properties, including high strength and low density, along with excellent corrosion resistance and biocompatibility [1-3,7-14].

Adequate corrosion resistance of these alloys is a basic requirement for the performance of dental implant rehabilitations. The oral medium contains potentially corrosive substances such as acids, chlorides, and fluorides. Fluoride ($F^-$) may be present in the oral medium at various concentrations and pH values. The use of fluoride is an excellent alternative to promote the prevention of caries and treatment of sensitivity after using dental bleaching agents. The fluoride interacts with the mineralised tissue of the tooth, the hydroxyapatite, forming fluorapatite [5,7,8,10,11,15].

Fluoride compounds used in oral applications are commonly found in association with sodium or in the phosphate acidified solution. The concentrations in the oral formulations are commonly within the range of 100–20000 ppm for fluorine ions ($F^-$) or NaF from 0.1 to 2% (wt/v). Fluoride ions are frequently available through food ingestion, drinking-water (0.7–1.0 ppm), or during topical application of toothpastes (1000–1500 ppm). Other sources of fluoride include daily mouthwashes (227 ppm), weekly mouthwashes (905–2270 ppm), professional dental gel/mousse (9050 ppm), professional acidified gel (12300 ppm), and professional fluoride varnish (22700 ppm) [8,10,11,13,16]. The use of acidulated gel can lead to the formation of calcium fluoride on the dental surface, which reduces the dental demineralisation process [3,8,11,13,16].

Titanium alloys exhibit high chemical reactivity in the presence of fluoride ions and low pH, as found in the oral environment [3,7,9-11,15,16]. Thus, an increase in fluoride concentration leads to the reduction of the corrosion resistance of Ti-based alloys. The pH can influence the corrosion resistance in combination with the fluoride concentration. Changes in the oral pH can be promoted by food, drink, mouthwashes, gel applications, biofilm metabolites, and inflammatory processes [2,3,5,8,10,11,13,14]. The acidification in the presence of fluorine can alter the surface roughness of the alloys because of the dissolution of titanium alloys. Indeed, hydrofluoric acid (HF) dissolves the protective film layer on the surface of titanium alloys [3,7,10,13,16]. A solution commonly reported in the literature is artificial saliva [1-5,7-11,13,15,17,18,23]. However, when NaF is added to artificial saliva that contains calcium ions, because of the lower solubility product constant of $CaF_2$ ($K_{sp}$ = 3.9 x $10^{-11}$ at 25 °C), a supersaturated solution is produced, from where fluorite precipitates.

The dental implant and rehabilitation system may fail as a result of loss of screw tightening or fitting, resulting in a microgap formation. The microgap allows relative micromovements in the joint region of rehabilitation between the components, such as abutments, screw, and implant, during masticatory or parafunctional loads. A microgap in range of 0.1 to 10 μm seems to be enough to micromovement occurrence on the rehabilitation joint [24,25]. For anti-rotational systems the micromovement may be present in range between 1.5 to 9.4 μm [24,25].

Moreover, saliva infiltration in the microgap increases the corrosion of metallic parts, and micromovement wears the implant-abutment, screw-implant, and screw-abutment surfaces. Finally, the torque relaxation of the prostheses-screw leads to failure. Additionally, oral fluids containing fluoride, acid, and biofilm can infiltrate the microgap. The simultaneous action of corrosion and wear is known as tribocorrosion, which is an important degradation mechanism with synergism effects that occurs during the relative micromovements between dental prostheses and implants. The use of titanium alloys in the oral medium, in contact with other alloys at the same time, can drive galvanic effects in association with acid and fluoride, causing a significant increase in tribocorrosion [1-5,7,10,11,15,17-21]. When NiCr is coupled with titanium alloys in oral environments, galvanic interactions are expected, thus causing severe corrosion of the alloy acting as the anode of the galvanic couple. Moreover, an electrochemical current can flow through the implant connections to reach the perimplant tissue, leading to bone resorption and pain [2,3,6,18,23].

The low wear resistance of dental alloys may influence the biocompatibility, generating an inflammatory response in the perimplant tissue because of the presence of debris released by tribocorrosion, promoting bone resorption and even the loss of implant osseointegration [1-3,7,10-12,22,23]. Titanium alloys may exhibit lower wear resistance then other alloys employed as abutments [7].

In the present work, the alloys commonly used in dental implant restoration systems were investigated, as they can undergo corrosion processes in the same environment [2,3,5]. For NiCr, the corrosion resistance is in general lower than that exhibited by titanium alloys under ordinary conditions [2,3]. The corrosion resistance of the NiCr casting alloys has in general been investigated without taking into account the influence of wear [2,5,18]. Studies addressing the tribocorrosion process of NiCr for biomaterials applications, or even galvanic coupling associated with tribocorrosion, are rarely found in the literature.

The present work aims to assess the effect of fluoride on the corrosion and tribocorrosion resistance of two alloys used in rehabilitation with dental implants. Furthermore, the effect of galvanic interactions concomitant with wear was studied for an implant system in two solutions with different aggressiveness.

## 2. Materials and Methods
### 2.1 Materials and specimen preparation

The alloys used in this study were the extra-low interstitial titanium alloy Ti6Al4V ELI (ASTM F136/ISO 5832-3, ACNIS, Brazil) and commercial dental NiCr used for casting (Neocast V, Dental Alloys, USA). The elemental composition of the alloys provided by the manufacturers is shown in Table 1. The casting conditions were the same as those used for manufacturing of dental supported implants; the procedure used was oxygen-gas flame melting and injection of the melted alloy into the mould by centrifugation in a dental prosthesis laboratory. The mould was made from a wax pattern embedded in a cast ring and covered by coating. The wax pattern was removed at a temperature of 850 °C. The alloy was heated within the melting range of 1150–1227 °C by a flame with oxygen and centrifuged for injection into the cast ring through feed channels. The melting process used an oxygen-gas flame followed by liquid injection into the mould by centrifugation. After solidification, the specimen was sandblasted with aluminium oxide to clean the surface.

*Table 1. Chemical composition of alloys (% wt).*

| Ti6Al4V (ASTM F136) | | | | NiCr (Neocast V) | |
|---|---|---|---|---|---|
| Al | 6.04 | C | 0.034 | Ni | 70–77 |
| V | 4.115 | H | 0.0028 | Cr | 11–14 |
| N | 0.004 | O | 0.105 | Mo | 8–10 |
| Fe | 0.185 | Ti | Balance | Al+Co+Ti | < 5 |

### 2.2 Tribocorrosion tests

For tribocorrosion tests, working electrodes with 25-mm diameter disks were progressively grounded until 600 mesh. A second electrode, in a bar shape, with an exposed area of 161 mm$^2$, was placed 5 mm vertically above the disks to simulate the galvanic effect. Both materials were used as disk and bar, as a single electrode (disk) or electrically connected (disk and bar). The area ratio of the disk to the second electrode (bar) was 3:1. The sample surfaces were then washed with distilled water, ultrasonically cleaned in isopropyl alcohol, and dried with warm air. All samples were stored in a drying chamber with silica gel for 24 h before the experiments. The second sample was connected to the disk under wear via a potentiostat to emulate the galvanic interaction. The counter-body used in contact was an alumina ball - ABNT ISO 3290-2 standard - with a 4.0-mm diameter (Só Esferas, Brazil). For each experiment, a new sphere surface was used.

The tribocorrosion test was performed in an electrochemical cell containing 100 mL of the selected solution attached to a rotational ball-on-plate tribometer, with the specimen surface facing upward against the counter-body alumina sphere. Figure 1 presents the scheme of the experimental setup used. An electrochemical cell configuration for control and measurement was used, with a silver/silver chloride (Ag/AgCl) reference electrode, and a working electrode (WE1 in Figure 1). The choice for sliding ball-on-plate was based on previous studies of Borrás et al. [26], Dimah et al. [27], Lepicka et al [28], Licausi et al. [29] and Licausi et al. [30] for evaluation dental implants, being a proper method to mimetic dental implant-abutment contact compared to the use of real implants. The electrochemical cell was machined in polymer, and an O-ring was attached to the working electrode to prevent solution leakage. A constant rotational frequency was imposed through a motor shaft attached to the holder. The electrochemical cell was connected to the potentiostat, while the alumina ball was kept fixed and connected to the tribometer. The sample holder allowed electrical contact between the working electrode with the potentiostat for single evaluation, or coupled for galvanic interaction evaluation.

Tribocorrosion tests were performed with both configurations, *i.e.*, single alloy (Ti6Al4V or NiCr) and couples formed by short-circuited electrodes. For the coupled materials condition, the alloy that slides against the counter-body was termed WE1. The electrochemical measurement performed during tribocorrosion tests was a total open circuit potential (OCP) of 9000 s before starting the sliding, 3600 s during sliding, and 1800 s after sliding. The imposed wear conditions were as follows: rotational frequency 1.0 Hz, 2.0 N normal contact load, corresponding to initial maximum Hertzian pressure of 814.5 MPa for NiCr and 998.7 MPa for Ti6Al4V. The initial contact pressure for Ti6Al4V was in accordance with others studies [26-29] and with the range of 333 MPa to 12 GPa as mentioned by Revathi et al. [31] for biomedical applications. The tests were carried out at room temperature (approximately 25 °C) using a Metrohm μAutolab 3AUT71295 digital potentiostat controlled by Autolab Nova 1.11 software. The friction load during the sliding was recorded through a load cell/LabView® interface program, which also allowed monitoring of the rotation speed and the start and finish of sliding.

Two combinations of fluoride content [F$^-$] and pH were selected: a less aggressive combination (227 ppm, pH 5.5), and a more aggressive one (12300 ppm, pH 4.0). The less aggressive condition corresponds to daily mouthwashes with the average pH found in the oral medium. The most severe one corresponds to the professional application of fluoride gel with a pH value associated with inflammatory situations. Lactic acid was used for pH adjustment. The pH of the solutions was determined using a 780 pH Meter (Metrohm Autolab BV, Switzerland). The chemical compositions of solutions were 0.9% NaCl+227 ppm F$^-$ and 0.9% NaCl+12300 ppm F$^-$.

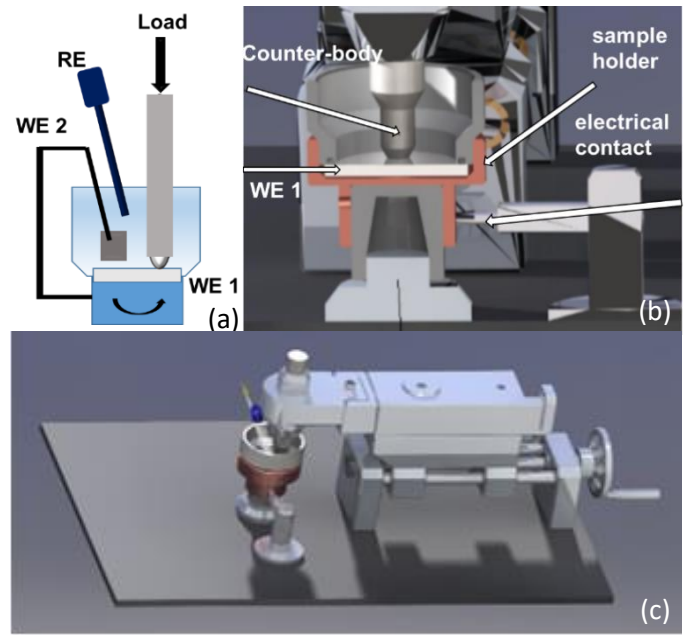

*Figure 1: Setup of tribometer adapted for galvanic corrosion measurements. (a) electrochemical cell, (b) internal view of cell, (c) tribometer.*

## 2.3 Surfaces characterization

After the tests, the worn surfaces were mapped using confocal microscope (LSM 800, Zeiss, Germany) in association with the software Conformap (Zeiss). The confocal microscope scanned the surface with a laser and obtained the 3D image of the track. These data were used to obtain, by polynomial integration, the wear volume loss of the track. Profiles perpendicular to the sliding direction were obtained from the worn tracks and used to reveal the deepest valleys.

## 2.4 Statistical Analysis

The experiments were performed in triplicate, and the representative result with the corresponding mean value and standard deviation were presented. The obtained data were statistically analysed with 1-way ANOVA variance and T-student tests. The parameters considered were open-circuit potential (OCP) before, during and after sliding; coefficient of friction (μ), amplitude of coefficient of friction (Δμ) and wear rate loss. The considered significance of statistical test for both analyses was $p < 0.05$.

## 3. Results and Discussion

The fluoride concentrations used here were 227 and 12300 ppm, combined with adjusted pH 5.5, representing a normal condition, or pH 4.0, representing conditions with acid biofilm metabolites or even inflammation, respectively [1-3,5,7,8,10,11,13,14,15-18,20,23]. The pH necessary to drastically deteriorate the corrosion resistance of titanium depends on fluoride concentration $[F^-]$, as shown in Eq. (1) developed by Nakagawa et al. [13], taking into account the polarization curve data without mechanical effort. In the present work, the selection of the environments with moderate and aggressive corrosion aspects in relation to titanium was based on Eq. (1). Thus, the moderate and aggressive solutions used on tribocorrosion tests were 227 ppm $F^-$ at pH 5.5 and 12300 ppm at pH 4.0, respectively. Moreover, the effect of mechanical wear was evaluated for these solutions with different aggressiveness to titanium alloy as well as the galvanic influence when coupled to NiCr.

$$pH < 1.49 \log[F^-] + 0.422 \quad (1)$$

Wear track profiles obtained with confocal laser analysis, measured perpendicularly to the sliding direction, are shown in Figure 2 for isolated Ti alloy and for coupled to NiCr, Figure 3 for isolated NiCr alloy and for coupled to Ti, in both solutions. The corresponding wear volume loss obtained from confocal analysis was converted to wear rate that is depicted in Figure 4. Clearly, the more aggressive solution, high fluoride content and lo pH, favours the higher volume loss of Ti measured as the profile of worn track regardless the galvanic couple with NiCr.

Isolated and coupled Ti alloy wear profiles exhibited deeper and rougher tracks, indicating the more severe degradation of this material in comparison with single and coupled NiCr. The deepest valley values were used to estimate the statistical significance. Under the same test conditions, single Ti alloy wear profiles were deeper in 12300 ppm/pH 4.0 than in 227 ppm/pH 5.5 ($p<0.05$). However, the values obtained for the couple Ti6Al4V/NiCr exhibited lower depth than single Ti6Al4V in both environments. The inverse couple, NiCr/Ti6Al4V exhibited higher penetration depth when compared to single NiCr in the more aggressive solution. This finding indicates a galvanic effect of Ti alloy in fluoride content. The wear volume loss obtained was converted to wear rate with Eq. (2), used by Lepicka et al. [28]:

$$Wear\ rate\ \left(\frac{mm^3}{N.m}\right) = \frac{Wear\ volume\ loss\ (mm^3)}{Apllied\ load\ (N)\ .\ Sliding\ distance\ (m)} \quad (2)$$

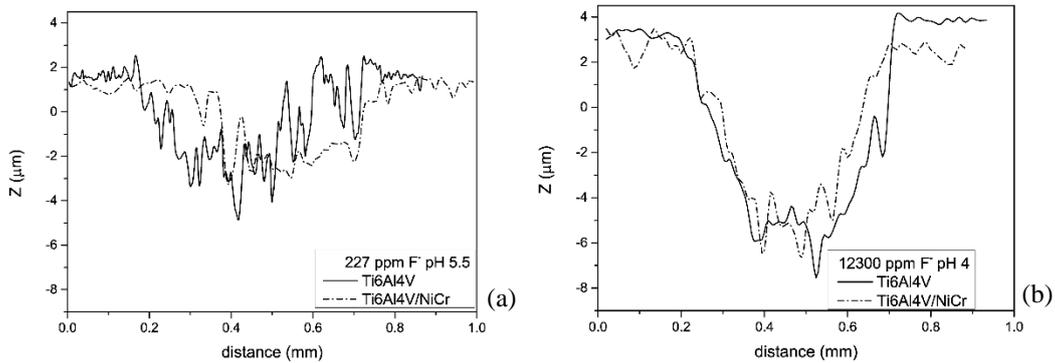

*Figure 2: Wear track profiles obtained for Ti6Al4V single alloy and in couple combinations of Ti6Al4V/NiCr in 227 pm at pH 5.5 (a) and 12300 ppm at pH 4.0 (b). Worn materials – Ti alloy.*

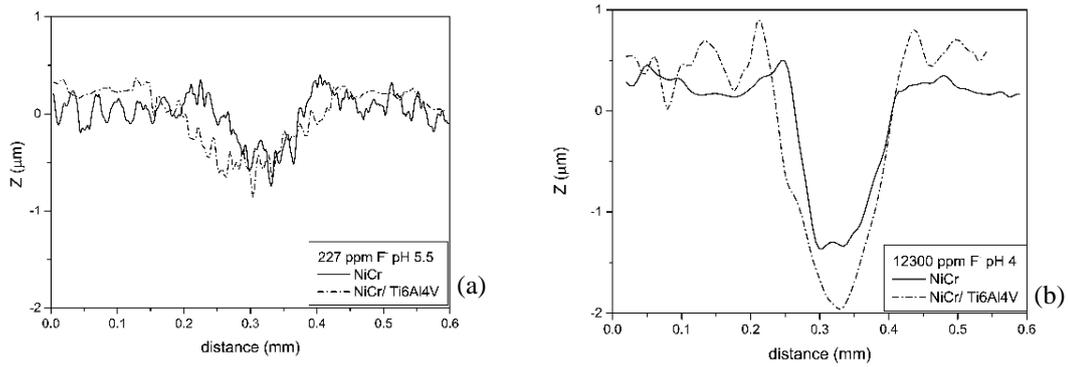

*Figure 3: Wear track profiles obtained for NiCr single alloy and in coupled combinations of NiCr/ Ti6Al4V in 227 ppm and pH 5.5 (a) and 12300 and pH 4.0 (b). Worn materials – NiCr alloy.*

The specific wear rate is depicted in Figure 4. Titanium alloy presented an increase in the wear rate with increasing fluoride concentration, as found by Sivakumar et al. [33]. Comparing the alloys tested under single and coupled configurations, significant differences in wear volume losses were observed in both solutions (p<0.05). Titanium alloy, either singly or coupled with NiCr, exhibited a specific wear rate *circa* five times that of the NiCr alloy (p<0.05). Furthermore, the NiCr wear rate was not strongly affected by the environment or galvanic interaction, with a slight increase in wear. However, Ti alloy exhibited significant variations between solutions (p < 0.05) and with and without the effect of galvanic interaction. This effect of galvanic interaction was beneficial for Ti because it decreases the volume loss.

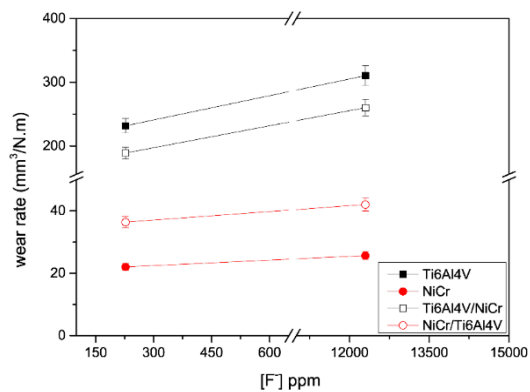

*Figure 4: Specific wear rate obtained for Ti6Al4V, Ti6Al4V/NiCr, NiCr, and NiCr/Ti6Al4V in 227 ppm/pH 5.5 and 12300 ppm/pH 4.0.*

The evolution of OCP before, during, and after sliding is shown in Figure 5, and the evolution of the coefficient of friction (µ) during sliding is given in Figure 6. NiCr has a higher potential than Ti alloy without friction. The friction provokes a potential decay, the corresponding parameters of which are shown in Tables 2 and 3.

Before sliding, the OCP of single NiCr and coupled NiCr/Ti6Al4V stabilised at similar values in both solutions, with $p>0.05$. For single Ti6Al4V and coupled Ti6Al4V/NiCr in 227 ppm/pH 5.5, before sliding, OCP also stabilised at similar values ($p>0.05$). However, for 12300 ppm/pH 4.0, the OCP of single Ti6Al4V and Ti6Al4V/NiCr were more unstable, with a difference in potential that could be associated with galvanic interaction ($p<0.05$). The single Ti6Al4V in 12300 ppm/pH 4.0 presented a continuous decrease in potential, with a tendency to stabilise within the thermodynamic corrosion domain of Ti. When coupled to NiCr, the Ti alloy in this solution presented a shift to higher OCP values, although with a continuous decrease in the potential over time. It is important to point out that the OCP of couples tends to stabilise within the thermodynamic passivation domain, as shown in Figure 5.

At the start of sliding, in all tested conditions, an immediate decay of the potential to lower values was observed because of the severe mechanical removal of the films formed at the metallic surface [7,13]. For NiCr alloy in 227 ppm/pH 5.5, single and coupled, lower potential values were maintained throughout the sliding process, recovering the initial values when the sliding was interrupted ($p>0.05$). This behaviour is associated with the fact that in the corrosive solution, the Ti alloy remains passive, as pointed out in Figure 5, without influencing the corrosion process of NiCr when associated therewith. In 12300 ppm/pH 4.0, single NiCr maintained lower values of potential during the sliding, while the coupled NiCr/Ti6Al4V presented a decreasing potential during sliding, but with no significant differences between the initial and final values until sliding stopped ($p>0.05$). This behaviour can be associated with titanium, which in this solution presents a tendency to corrode, making the coupled potential lower during sliding. For all conditions, at the end of sliding, the potential presented values close to the initial value, indicating the repassivation ability of NiCr. However, after sliding in 12300 ppm/pH 4.0, the coupled NiCr/Ti6Al4V presented a tendency to decrease potential, as a consequence of the potential variation of the Ti alloy towards the corrosion domain in this solution.

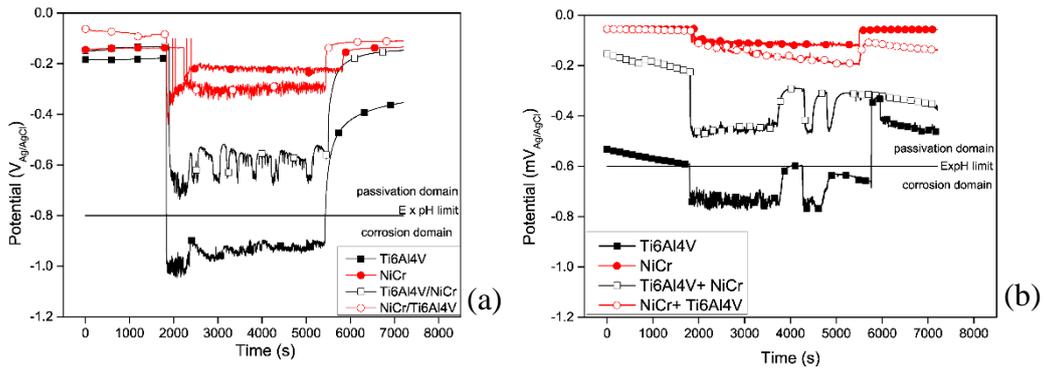

*Figure 5: OCP measurements during tribocorrosion tests for single and coupled combinations of NiCr and Ti alloy in 227 ppm/pH 5.5 (a), and 12300 ppm/pH 4.0 (b).*

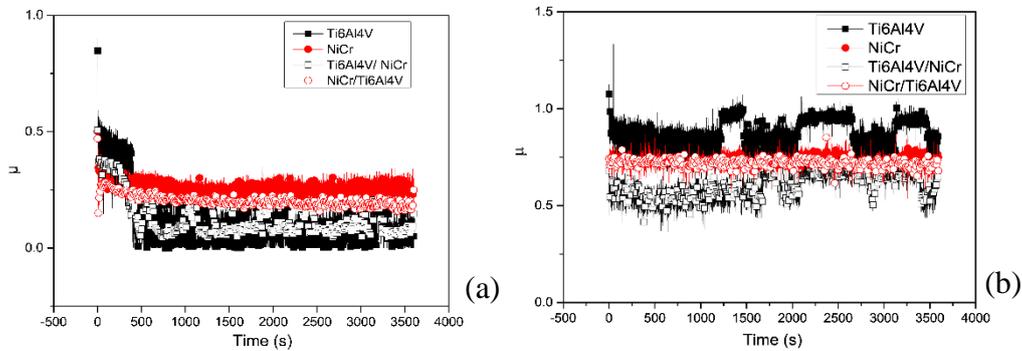

*Figure 6: Coefficient of friction measurements during sliding from tribocorrosion tests of single and coupled combinations of NiCr and Ti alloy in 227 ppm/pH 5.5 (a) and 12300 ppm/pH 4.0 (b).*

*Table 2: Parameters obtained from tribocorrosion tests in 227 ppm/pH 5.5.*

| Alloy | Initial potential (V) | Initial sliding potential (V) | Final sliding potential (V) | Final potential (V) | $\mu$ | $\Delta\mu$ |
|---|---|---|---|---|---|---|
| Ti6Al4V | -0.184 ±0.018 | -1.027 ±0.089 | -0.910 ±0.10 | -0.368 ±0.077 | 0.14 ±0.022 | 0.22 ±0.013 |
| NiCr | -0.063 ±0.082 | -0.434 ±0.047 | -0.296 ±0.007 | -0.110 ±0.049 | 0.26 ±0.015 | 0.10 ±0.013 |
| Ti6Al4V/NiCr | -0.148 ±0.031 | -0.722 ±0.007 | -0.564 ±0.028 | -0.148 ±0.033 | 0.15 ±0.013 | 0.13 ±0.054 |
| NiCr/Ti6Al4V | -0.144 ±0.005 | -0.265 ±0.042 | -0.227 ±0.002 | -0.134 ±0.031 | 0.20 ±0.023 | 0.08 ±0.006 |

*Table 3: Parameters obtained from tribocorrosion tests in 12300 ppm/pH 4.0*

| Alloy | Initial potential (V) | Initial sliding potential (V) | Final sliding potential (V) | Final potential (V) | µ | Δµ |
|---|---|---|---|---|---|---|
| Ti6Al4V | -0.400 ±0.018 | -0.715 ±0.10 | -0.634 ±0.065 | -0.486 ±0.041 | 0.87 ±0.033 | 0.21 ±0.032 |
| NiCr | -0.057 ±0.002 | -0.129 ±0.023 | -0.138 ±0.026 | -0.063 ±0.008 | 0.79 ±0.035 | 0.12 ±0.025 |
| Ti6Al4V/NiCr | -0.168 ±0.020 | -0.472 ±0.017 | -0.454 ±0.007 | -0.376 ±0.014 | 0.73 ±0.017 | 0.13 ±0.060 |
| NiCr/Ti6Al4V | -0.074 ±0.029 | -0.191 ±0.11 | -0.283 ±0.013 | -0.206 ±0.090 | 0.70 ±0.028 | 0.10 ±0.035 |

Cruz et al. [17] mentioned that higher OCP during sliding indicates a lower tribocorrosion tendency for materials. Based on this criterion, the NiCr alloy presented a lower susceptibility to wear-induced corrosion than Ti, as indicated by the results of tribocorrosion tests obtained in the present work. Lower OCP after sliding tests for single Ti6Al4V was also reported by Souza et al. [7] in artificial saliva with 12300 ppm/pH 5.5, resulting from continuous active dissolution of the alloy.

The decrease in OCP for single Ti6Al4V during sliding in both solutions presented relevant differences in the potential values achieved, as identified by $p<0.05$. Nevertheless, this difference was not important for the isolated material, as in both conditions, the alloy under sliding was within the corrosion domain, as illustrated in Figure 5. However, when coupled to NiCr, although the coupled Ti6Al4V/NiCr still shows a more pronounced decrease in OCP during sliding, the presence of NiCr in the galvanic couple moves the potential to the passive domain of Ti. Despite the well-known corrosion susceptibility of Ti alloys in combination with high fluoride concentrations and lower pH, [1-3,7-11,13-18,20,23], when coupled to NiCr, the titanium alloy remained passive, even in the presence of a highly aggressive combination of fluoride and pH.

For single Ti6Al4V in 12300 ppm/pH 4.0, the most aggressive solution, the alloy remains within the thermodynamic corrosion domain from the start. The potential decay observed after the onset of sliding was not only associated with the mechanical damage of a protective film, but also with the enhancement of active dissolution on the newly exposed surface, leading to a lower final potential [8]. In this case, the mass loss occurred on the entire surface and cannot be associated only with wear loss.

During sliding, it is possible to observe fluctuations in the potential, which behaviour could be associated with the simultaneous processes of passivation on the exposed rubbed area, while the film damage progresses on the next area, as also reported by Davoren et al. [21]. For Ti alloys, these fluctuations decreased with increasing fluoride concentration because of the influence of fluoride ions on passive film formation. Another aspect is related to the negative potential shift during the initial sliding period, as cited by Sivakumar et al. [33], with an increase in fluoride concentration.

The literature mentions that weight loss of Ti alloys under tribocorrosion increases with higher fluoride concentrations. In the present work, 3D reconstruction of the total wear track allowed the limitation of using an analytical balance to determine the wear volume loss to be overcome. Espallargas et al. [22] also obtained volume loss corresponding to a worn surface, as did Wimmer et al. [32]. However, Espallargas et al. [22] obtained the loss through multiplication of the length by the average cross-section profile of the wear track, and Wimmer et al. [32] used an interferometer. In our study, the total volume loss of the wear track was obtained through a polynomial approximation as a function of the surface obtained by the Conformap software.

The potential drop for the Ti6Al4V alloy associated with the sliding has a correlation with the wear volume loss, as reported by Swaminathan and Gilbert [19]. From the results obtained in the present work, the Ti6Al4V alloy presented more severe degradation during tribocorrosion than the coupled Ti6Al4V/NiCr. A correlation with OCP decay induced by sliding has to also take into account the persistence of the potential under sliding in the passive or active dissolution domains. This analysis explains the reduction of the wear volume loss of the titanium alloy when coupled to NiCr. During the sliding process, a galvanic interaction is established between the worn area and the remaining surface, depending on the potential difference between the regions. More intense galvanic effects during sliding affect the wear volume loss because of the synergism between corrosion and wear [7,10,11].

Figures 7 and 8 show the wear tracks of alloys after sliding in 227 ppm/pH 5.5 and 12300 ppm/pH 4.0, respectively. NiCr exhibited a track aligned with the sliding direction, with marks corresponding to the position where the alumina ball was in contact with the sliding surface. No differences in the wear track were observed for the isolated NiCr and coupled NiCr to Ti6Al4V in both solutions. For the Ti alloy, the wear track showed areas associated with plastic deformation and detachment of the surface for isolated Ti6Al4V and coupled Ti6Al4V to NiCr in 227 ppm/pH 5.5, although it was less intense for the couple. In 12300 ppm/pH 4.0, isolated Ti6Al4V did not present persistent areas associated with plastic deformation because of the intense dissolution, while for the Ti6Al4V coupled to NiCr, it was possible to observe areas associated with plastic deformation and detachment of the material; however, these area were smaller than they were in 227 ppm F$^-$ pH 5.5. The intense dissolution of isolated Ti6Al4V in 12300 ppm/pH4.0 was identified by Barros et al. [34]. The analysis carried out by the authors through SEM and ICP-MS techniques allowed to characterize the surface morphology and ions released and the maintenance on a passive state of coupled Ti6Al4V.

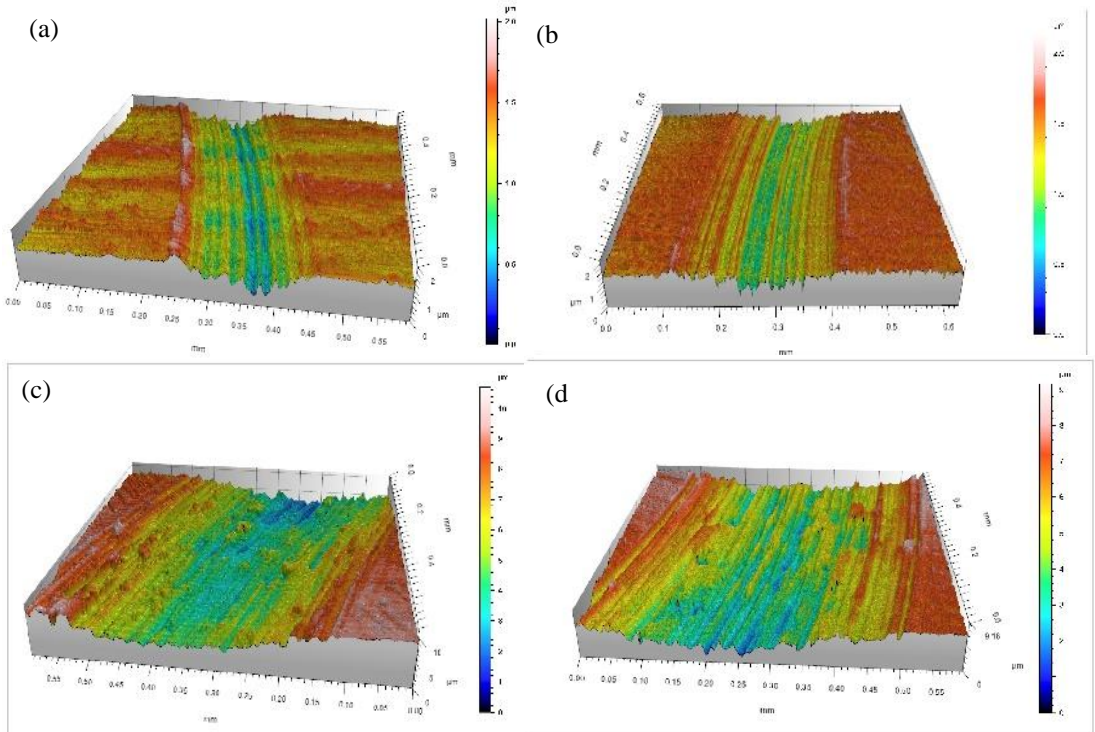

*Figure 7: 3D image of a section of worn track for single NiCr (a), coupled NiCr/Ti6Al4V (b), single Ti6Al4V (c), and coupled Ti6Al4V/NiCr (d) after sliding tribocorrosion tests in 227 ppm fluoride and pH 5.5.*

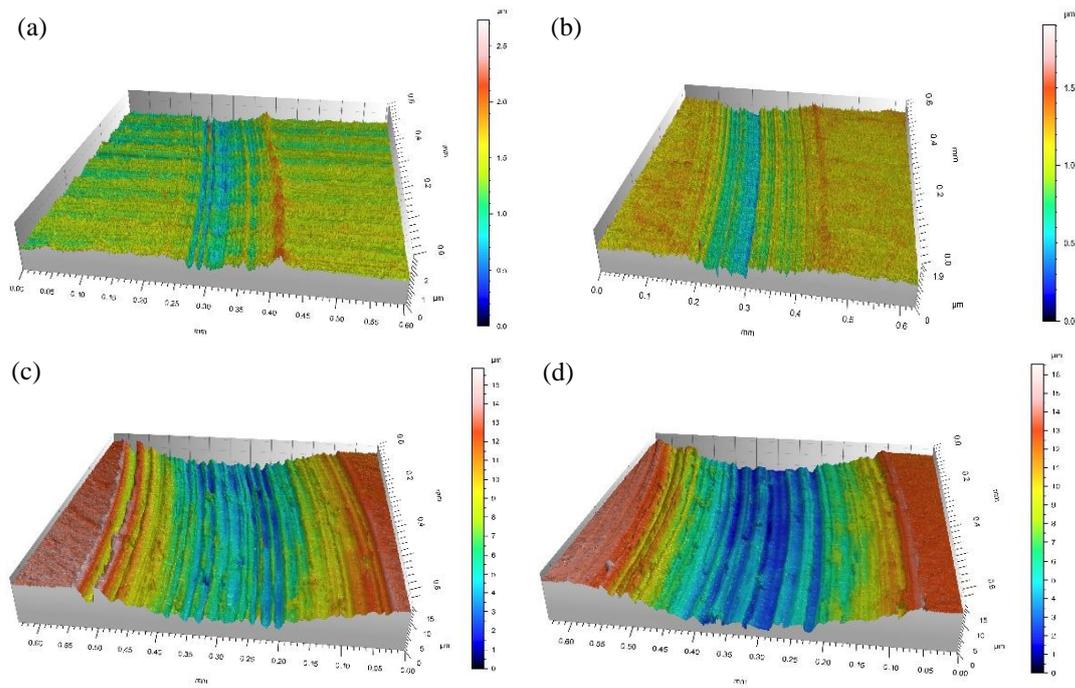

*Figure 8: 3D image of a section of worn track for single NiCr (a), coupled NiCr/Ti6Al4V (b), single Ti6Al4V (c), and coupled Ti6Al4V/NiCr (d) after sliding tribocorrosion tests in 12300 ppm fluoride and pH 4.0.*

The evolution of the coefficient of friction is shown in Figures 5 and 6, with the respective averages given in Tables 2 and 3. The coefficient of friction calculated during the sliding steps exhibited a quick initial increase, corresponding to the destruction of the passive film, as mentioned in the literature [1], or to the removal of the first layer of material and initiation of the wear track. The subsequent evolution of the coefficient of friction suggests that after the initial step of film destruction, the alumina ball contacted the ground-sliding surface and the quasi-stationary process started. Results obtained with the NiCr alloy under sliding did not present significant variation for the coefficient of friction between the isolated and coupled conditions, for each solution (p>0.05). However, the increase in fluoride concentration increased the coefficient of friction (p<0.05), without affecting its amplitude (p>0.05), for both alloys. Ti alloy sliding surfaces presented a decay in the coefficient of friction when under galvanic interaction with NiCr, for both solutions (p<0.05), as well for the amplitude of the coefficient of friction (p<0.05). Figure 9 presents the observed correlation that can be established during the sliding between OCP and the coefficient of friction for isolated and coupled Ti6Al4V. These variations can be associated with the incidence of plastic deformation, creation of new wear track lines, and debris acting as a third body on the wear tracks [7,11].

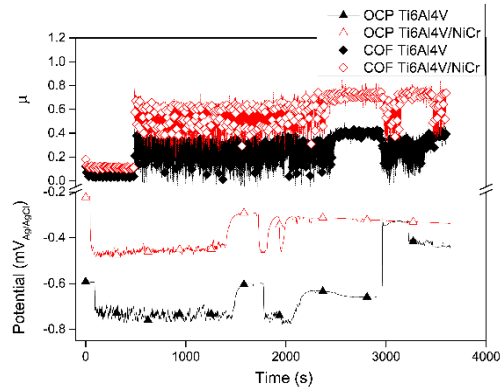

*Figure 9: Correlation between coefficient of friction and OCP of Ti6Al4V during sliding tribocorrosion tests in 12300 ppm/pH 4.0 solution.*

Lower values for the coefficient of friction have been reported in higher fluoride concentration solutions without considering the relevant influence of pH [7,11]. The difference found in comparison with our study was that the anti-wear action was attributed to the formation and deposition of $CaF_2$, as saliva plus NaF solution was the environment used in previous studies [17]. The continuous salt precipitation reduces friction and wear effects under lower shear stress, and can explain the decrease in the coefficient of friction with increasing fluoride concentration. Cruz et al. [17] mentioned that a minimum concentration of 100 ppm/pH 5.0 is required to promote the spontaneous formation of $CaF_2$ in artificial saliva. In our study, because of the possible formation of this precipitate, which can influence the behaviour, we chose a NaCl-based solution. Without the protection provided by the $CaF_2$ precipitates, the Ti alloy presents a susceptibility to tribocorrosion, enhanced by the synergism between wear and high fluoride concentration and low pH. The lubricant ability reported in the literature associated with $CaF_2$ gradually decreases the coefficient of friction until the alumina ball state on the rubbing surface due the continuous salt formation and deposition.

The results revealed that the Ti alloy in galvanic couple with NiCr presented higher resistance than isolated Ti6Al4V. These results are coherent with the analysis of galvanic tribocorrosion presented by Swaminathan and Gilbert [19]. In their study, the authors evaluated the pairs Ti6Al4V/Ti6Al4V, CoCrMo/Ti6Al4V and CoCrMo/CoCrMo in phosphate–buffered saline solution (PBS). As in our study, with NiCr instead of CoCrMo, the couple with the mixed alloys presented no significant differences to the CoCrMo, and with less wear volume loss, potential and coefficient of friction than for the Ti6Al4V, due the galvanic interaction.

The volume loss due to tribocorrosion on implant systems has a clinical implication. The loss of seal between the abutment prosthesis and implant allows the formation of a microgap where infiltration of microorganisms is possible. Further, the increase on roughness associated to the corrosion and wear in this microgap region favours the adhesion of biofilms. This scenario, associated to the harmful action of fluoride on titanium, under continuous masticatory loads, could promote torque relaxation of the screw abutment, inflammatory process, local acidification, bone loss and risks to the osseointegration. These factors can be considered as significant causes of dental implant rehabilitation failure. Although there is a consensus that the presence of different metallic materials could generate a galvanic cell and more intense corrosion, it was concluded that the galvanic couple formed between Ti alloy and NiCr in fluoride solutions reduces the tribocorrosion severity of Ti alloy when compared to the use of the single Ti alloy.

## 4. Conclusions

Experiments on single electrode and galvanic coupling demonstrated the NiCr presented higher tribocorrosion resistance than Ti6Al4V in media containing chloride and fluoride. The higher fluoride concentration reduced the tribocorrosion resistance of Ti6Al4V, detected by the increases of the wear rate, the potential drop and the coefficient of friction during the sliding, while the casting NiCr alloy was not affected.

The galvanic interaction between the Ti6Al4V and NiCr alloys in solutions with fluoride concentration and pH variations is possible and allowed the stabilization at potential where titanium alloy can passivate. Even when Ti6Al4V alloy was under the most aggressive combination of fluoride and acidity, and coupled with NiCr, the titanium present in Ti6Al4V altered from the corrosion domain to passive domain before, during and after sliding. Consequently, the tribocorrosion susceptibility of Ti6Al4V was reduced when connected to NiCr as consequence of the galvanic interaction which changed the coupled electrochemical potential. In summary, for a total cathode/anode area ratio equal to 1:3, the wear rate increases in the sequence NiCr < NiCr-Ti6Al4V < Ti6Al4V-NiCr < Ti6Al4V in 0.9% NaCl with fluoride. Hence, the galvanic effect increases the wear rate of NiCr and reduces for Ti6Al4V.

## 5. Acknowledgments

The authors acknowledge the support by CNPq, Faperj and Fundação Coppetec/Brazil. This investigation was supported in part by the Coordenação de Aperfeiçoamento de Pessoal de Nível Superior - Brasil (CAPES) - Finance Code 001.